\documentclass[11pt]{article}
\usepackage{latexsym,subfigure, url}

\usepackage{t1enc}
\usepackage[english]{babel}
\usepackage{amsmath, amsfonts, amsthm, amsxtra, amssymb}
\usepackage{color}
\usepackage{times}
\usepackage{graphicx}

\newcommand{\argmin}{\mbox{argmin}}
\newcommand{\piRW}{\pi_{\text{RW}}}
\newcommand{\piID}{\pi_{\text{ID}}}
\newcommand{\email}[1]{{\tt #1}}

\title{Efficient Algorithms for Sampling and Clustering of Large
Nonuniform Networks
}

\author{
Pekka Orponen\thanks{Research supported by the Academy of Finland,
grant 204156. Part of the research was done during this author's
visit to the Santa Fe Institute in March--April 2004.
E-mail: {\email{orponen@tcs.hut.fi}}}
\qquad
Satu Elisa Schaeffer\thanks{Research supported by the Academy of
Finland, grant 206235, and the Nokia Foundation.
E-mail: {\email{satu@tcs.hut.fi}}}\\
Laboratory for Theoretical Computer Science, P.O. Box 5400 \\
FI-02015 Helsinki University of Technology, Finland
}

\date{}

\begin{document}

\maketitle

\begin{abstract}
We propose efficient algorithms for two key tasks in
the analysis of large nonuniform networks: uniform node
sampling and cluster detection.
Our sampling technique is based on augmenting a simple, but
slowly mixing uniform MCMC sampler with a regular random
walk in order to speed up its convergence; however the
combined MCMC chain is then only sampled when it is in its 
``uniform sampling'' mode.
Our clustering algorithm determines the relevant neighbourhood
of a given node $u$ in the network by first estimating the Fiedler
vector of a Dirichlet matrix with $u$ fixed at zero potential,
and then finding the neighbourhood of $u$ that yields a minimal
weighted Cheeger ratio, where the edge weights are determined by
differences in the estimated node potentials.
Both of our algorithms are based on local computations, i.e.\
operations on the full adjacency matrix of the network are not used. 
The algorithms are evaluated experimentally using three types of
nonuniform networks: Dorogovtsev-Goltsev-Mendes
``pseudofractal~graphs'', scientific collaboration networks, 
and randomised ``caveman graphs''.

\end{abstract}

\section{Introduction}   \label{sec:introduction}

Two key tasks in the analysis of large natural networks, such as
communication networks and social networks, are obtaining a {\em
uniform sample} of nodes in the network, and determining the densely
interconnected {\em clusters} of nodes. Uniform sampling is important
e.g.\ for the purpose of estimating basic network characteristics such
as the degree distribution, average path length, and clustering
coefficient; it is, however, nontrivial to obtain a truly uniform
random sample of nodes from a large, practically unobtainable network
such as the WWW~\cite{HHMN00}. In this paper, we suggest an efficient
approach for uniform sampling of undirected nonuniform graphs, using a
construction that combines two types of random walks to produce one
that mixes rapidly and still converges to the uniform distribution
over the set of nodes.

We also discuss the problem of clustering nonuniform networks, i.e.\
the recognition of subgraphs where the nodes have relatively many
edges among themselves and relatively few edges connecting them to the
rest of the graph~\cite{KaVV04}. For large nonuniform networks, an
effective clustering algorithm should scale at most linearly in the
size of the graph, and for many applications, a method for determining
the local cluster of a given source node will suffice, rather than a
complete clustering of the entire graph. In this paper, we use
approximate Fiedler vectors to determine potentials around a given
source node, and then use the potentials to stochastically select an
appropriate local cluster.

In Section 2, we present the MCMC construction for uniform sampling,
and in Section 3 discuss experiments performed with the
method. Section 4 discusses local clustering with Fiedler
vectors. Finally, Section 5 summarises the work and addresses
directions for further research.

\section{An Efficient MCMC Method for Uniform Sampling}
\label{sec:sampling_method}

Let $G = (V,E)$ be a connected symmetric simple graph
with $n$ nodes and $m$ edges.
We denote the {\em neighbourhood} of node $i \in V$ by
$\Gamma(i) = \{j \in V \mid (i,j) \in E\}$,
and the {\em degree} of $i$ by $\deg(i) = |\Gamma(i)|$.
It is well known
(and easy to verify) that the {\em regular} random walk on $G$,
with transition probabilities
\begin{equation}   \label{eq:regular}
p_{ij} = \left\{
   \begin{array}{ll}
      \displaystyle{\frac{1}{\deg(i)}},  & \quad \text{if } j \in \Gamma(i), \\
      0,                                 & \quad \text{otherwise,}
   \end{array} \right.
\end{equation}
satisfies the {\em detailed balance} conditions
\begin{equation}   \label{eq:detbal}
\forall i, \, j \in V : \pi_i \cdot p_{ij} = \pi_j \cdot p_{ji}
\end{equation}
with respect to the distribution $\pi_i = \deg(i)/2m$, and hence this
distribution, which we denote by $\piRW$, is stationary w.r.t.\ the
walk~\cite{Brem99,Hagg02}.  If $G$ is non-bipartite, then $\piRW$ is
the unique equilibrium distribution.  The chain~(\ref{eq:regular})
mixes rapidly, but the probability of obtaining any given node $i$ as
a sample from it is proportional to the degree of $i$, and thus not
uniform unless $G$ is regular.

A straightforward approach to uniform sampling \cite{BBCF00}
would be to augment the nodes of $G$ with virtual
self-loops so as to make them all have the same degree
$d = \max_{i \in V} \deg(i)$. This method, however,
requires knowing the target degree $d$ ahead of time, and such
global information is typically not available in many of the
interesting applications.
Moreover, this process may create some convergence
anomalies in the case of highly nonuniform graphs $G$.
Another alternative~\cite{HHMN00,RPLG01} would be to postprocess
a sample obtained from walk~(\ref{eq:regular}) in order to
compensate for the bias in the stationary distribution $\piRW$.
Such postprocessing, however, requires some {\em a priori} information
on the number of burn-in steps needed before one can obtain
a representative sample from $\piRW$, and the burn-in time
again depends on the global structure of $G$.

We take a complementary approach, by starting from a somewhat
more slowly mixing random walk on $G$ with a provably uniform
stationary distribution, and then ``accelerate''~this walk by
coupling it together with the chain~(\ref{eq:regular}); however
we only sample the combined process when it is in the
``uniform sampling'' mode.

More precisely, we take as our starting point the following
{\em degree-balanced} random walk on $G$, where the transition
probabilities from node $i$ are inversely proportional to the
degree of the target node $j$:
\begin{equation}   \label{eq:balanced}
p_{ij} = \left\{
   \begin{array}{ll}
      \displaystyle\frac{1}{\deg(i) \cdot \deg(j)},
         & \text{ if } j \in \Gamma(i), \\
      1 - \displaystyle\sum_{j \in \Gamma(i)}
         \displaystyle\frac{1}{\deg(i) \cdot \deg(j)},
         & \text{ if } j = i, \\
      0, & \text{ otherwise.}
   \end{array} \right.
\end{equation}
It is simple to verify that the transition probabilities $p_{ij}$
given by~(\ref{eq:balanced}) satisfy the detailed balance
conditions  with respect to the uniform
distribution $\piID(i) = 1/n$, and hence $\piID$ is a stationary
distribution for this chain. (Note that in this case the equilibrium
distribution is unique for any $G$ with more than two nodes, since
any node $i$ with non-leaf neighbours has 
a self-loop probability of $p_{ii} > 0$.)

However, this degree-balanced walk avoids visiting the
high-degree nodes of a nonuniform graph, and so
mixes relatively poorly in the graphs of most interest to us.
A related problem is that the self-loop probabilities $p_{ii}$ are
rather large for nodes with many high-degree neighbours.\footnote{
These problems could be alleviated somewhat by using
the {\em Metropolis-Hastings} chain proposed in~\cite{BoDX04}, with
$p_{ij} = \min \{1/d_i, 1/d_j\}$ for $j \in \Gamma(i)$,
instead of our degree-balanced chain. However, as illustrated
in Figure~\ref{fig:dgmtvd} below, both chains have qualitatively
similar convergence behaviour, and the arithmetic of
coupling to the regular random walk is somewhat simpler for
the degree-balanced version.}

\begin{figure}
\begin{tabular}{cc}
\phantom{spacing} & 
\begin{picture}(0,0)%
\includegraphics{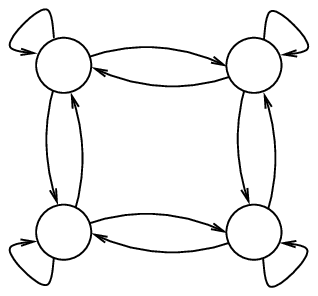}%
\end{picture}%
\setlength{\unitlength}{4144sp}%
\begingroup\makeatletter\ifx\SetFigFont\undefined
\def\x#1#2#3#4#5#6#7\relax{\def\x{#1#2#3#4#5#6}}%
\expandafter\x\fmtname xxxxxx\relax \def\y{splain}%
\ifx\x\y   
\gdef\SetFigFont#1#2#3{%
  \ifnum #1<17\tiny\else \ifnum #1<20\small\else
  \ifnum #1<24\normalsize\else \ifnum #1<29\large\else
  \ifnum #1<34\Large\else \ifnum #1<41\LARGE\else
     \huge\fi\fi\fi\fi\fi\fi
  \csname #3\endcsname}%
\else
\gdef\SetFigFont#1#2#3{\begingroup
  \count@#1\relax \ifnum 25<\count@\count@25\fi
  \def\x{\endgroup\@setsize\SetFigFont{#2pt}}%
  \expandafter\x
    \csname \romannumeral\the\count@ pt\expandafter\endcsname
    \csname @\romannumeral\the\count@ pt\endcsname
  \csname #3\endcsname}%
\fi
\fi\endgroup
\begin{picture}(3130,1394)(-10,-574)
\put(-10, 79){\makebox(0,0)[lb]{\smash{\SetFigFont{12}{14.4}{rm}{\color[rgb]{0,0,0}Sampling side}%
}}}
\put(3120, 78){\makebox(0,0)[lb]{\smash{\SetFigFont{12}{14.4}{rm}{\color[rgb]{0,0,0}Mixing side}%
}}}
\put(2437,-321){\makebox(0,0)[lb]{\smash{\SetFigFont{10}{12.0}{rm}{\color[rgb]{0,0,0}$j'$}%
}}}
\put(1554,-321){\makebox(0,0)[lb]{\smash{\SetFigFont{10}{12.0}{rm}{\color[rgb]{0,0,0}$j$}%
}}}
\put(1561,438){\makebox(0,0)[lb]{\smash{\SetFigFont{10}{12.0}{rm}{\color[rgb]{0,0,0}$i$}%
}}}
\put(2430,438){\makebox(0,0)[lb]{\smash{\SetFigFont{10}{12.0}{rm}{\color[rgb]{0,0,0}$i'$}%
}}}
\put(1215,693){\makebox(0,0)[lb]{\smash{\SetFigFont{9}{10.8}{rm}{\color[rgb]{0,0,0}$p_{ii}$}%
}}}
\put(2627, 84){\makebox(0,0)[lb]{\smash{\SetFigFont{9}{10.8}{rm}{\color[rgb]{0,0,0}$p_{j'i'}$}%
}}}
\put(2705,712){\makebox(0,0)[lb]{\smash{\SetFigFont{9}{10.8}{rm}{\color[rgb]{0,0,0}$p_{i'i'}$}%
}}}
\put(1175,-510){\makebox(0,0)[lb]{\smash{\SetFigFont{9}{10.8}{rm}{\color[rgb]{0,0,0}$p_{jj}$}%
}}}
\put(1953,641){\makebox(0,0)[lb]{\smash{\SetFigFont{9}{10.8}{rm}{\color[rgb]{0,0,0}$p_{ii'}$}%
}}}
\put(1920,-119){\makebox(0,0)[lb]{\smash{\SetFigFont{9}{10.8}{rm}{\color[rgb]{0,0,0}$p_{jj'}$}%
}}}
\put(1924,-489){\makebox(0,0)[lb]{\smash{\SetFigFont{9}{10.8}{rm}{\color[rgb]{0,0,0}$p_{j'j}$}%
}}}
\put(2698,-534){\makebox(0,0)[lb]{\smash{\SetFigFont{9}{10.8}{rm}{\color[rgb]{0,0,0}$p_{j'j'}$}%
}}}
\put(1287,107){\makebox(0,0)[lb]{\smash{\SetFigFont{9}{10.8}{rm}{\color[rgb]{0,0,0}$p_{ij}$}%
}}}
\put(1911,293){\makebox(0,0)[lb]{\smash{\SetFigFont{9}{10.8}{rm}{\color[rgb]{0,0,0}$p_{i'i}$}%
}}}
\put(1771,108){\makebox(0,0)[lb]{\smash{\SetFigFont{9}{10.8}{rm}{\color[rgb]{0,0,0}$p_{ji}$}%
}}}
\put(2124,107){\makebox(0,0)[lb]{\smash{\SetFigFont{9}{10.8}{rm}{\color[rgb]{0,0,0}$p_{i'j'}$}%
}}}
\end{picture}
\end{tabular}
\caption{A diagram of the mirror construction for two nodes $i$ and
$j$ on the sampling side and their mirror nodes $i'$ and $j'$ on
the mixing side.}
\label{fig:mirror}
\end{figure}
In order to construct a sampling method that produces uniformly
distributed samples but avoids the convergence problems of
chain~(\ref{eq:balanced}), we propose the following construction (cf.\
Figure~\ref{fig:mirror}): for each node $i \in V$ we create a ``mirror
node'' $i'$.  The original nodes $i \in V$ are called the ``sampling
side'' and the mirror nodes $i' \in V'$ are the ``mixing side'' of the
augmented graph ($|V| = |V'| = n)$. We continue to denote by $\deg(i)
= \deg(i')$ the degree of $i$ {\em in the original graph $G$}, i.e.,
ignoring the added edges that connect the two sides.

The transition probabilities on the sampling side follow those of the
degree-balanced random walk; on the mixing side, a regular random walk
is mimicked with minor modifications. The exact transition
probabilities are defined as follows: let $\epsilon$ be a parameter
satisfying $0 < \epsilon < p_{ii}$ for all $i \in V$ --- further
restrictions on $\epsilon$ are discussed later in this section.  Fix
all the sampling-to-mixing transition probabilities $p_{ii'}$ to
$\epsilon$. On the sampling side, subtract $\epsilon$ from each
$p_{ii}$ and give all other transition probabilities the values they
would have in the degree-balanced walk. On the mixing side, denote the
probability of moving back to the sampling side from nodes $i'$ by
$p_{i'i} = \epsilon'_i$. Let $\delta$ be a parameter (to be determined
later) such that $\delta \geq \epsilon'_i$ for all $i' \in V'$.  Add
to each node $i' \in V'$ a self-loop with transition probability
$p_{i'i'} = \delta - \epsilon'_i$, and divide the remaining
probability mass $1- \delta$ evenly among the neighbours of $i'$ as in
a regular random walk, i.e.\ assign $p_{i'j'} = (1 - \delta)
\frac{1}{\deg(i)}$ for each $j' \in \Gamma(i') \setminus \{i\}$.

We claim that the stationary distribution of such a
{\em combination walk} is a weighted combination
of the distributions $\pi_{\text{ID}}$ and $\pi_{\text{RW}}$, such
that an $\alpha$-fraction of the time the  chain is in a state
$i \in V$, and an $(1- \alpha)$-fraction of the time is spent within
$V'$:
\begin{equation}
\pi_C(x) = \left\{
   \begin{array}{ll}
      \alpha \cdot \piID(x) = \alpha \cdot \frac{1}{n},
         & \text{ if } x = i \in V, \\
      (1-\alpha) \cdot \piRW(x) = (1-\alpha) \cdot \frac{deg(i)}{2m},
         & \text{ if } x = i' \in V'. \\
   \end{array} \right.
\end{equation}
To verify the claim it suffices to check the detailed balance conditions
(\ref{eq:detbal}) for the above construction. There are three
cases to consider:
(i) transitions within $V$,
(ii) transitions within $V'$, and 
(iii) transitions between $V$ and $V'$.

The first two cases are essentially the same as those considered in
the settings of the balanced and regular random walks, respectively:
only some constant coefficients ($\alpha$, $(1-\alpha)$, $(1-\delta)$)
appear on both sides of the balance equations and cancel out.
This leaves us with the third type: here the requirement is that
any transitions between a node $i \in V$ and its mirror node $i' \in V'$
satisfy
$\pi_C (i) \cdot p_{ii'} = \pi_C(i') \cdot p_{i'i}$, i.e.\ that
\begin{equation}   \label{eq:combi_detbal}
\alpha \frac{1}{n} \cdot \epsilon = 
   (1 - \alpha) \frac{\deg(i)}{2m} \cdot \epsilon'_i,
   \quad \text{for all $i \in V$.}
\end{equation}
These equations can be satisfied by solving for the transition
probabilities $\epsilon'_i$, once values for the parameters
$\alpha$ and $\epsilon$ have been chosen:
\begin{equation}
\epsilon'_i = \frac{2 m \alpha \epsilon}{n (1 - \alpha) \deg(i)}
= \frac{2m}{n} \cdot \frac{\alpha}{(1 - \alpha)} \epsilon \deg(i)^{-1},
\end{equation}
where $\frac{2m}{n} = \bar{k}$ is the average degree of nodes in $G$.
As a probability, $\epsilon'_i$ must be at most one for all $i \in V$.
This yields an additional restriction on the parameter $\epsilon$:
\begin{equation}   \label{eq:resteps}
\epsilon 
   \leq \frac{n}{2m} \cdot \frac{1 - \alpha}{\alpha} \deg(i)
   = \frac{1}{\bar{k}} \cdot \frac{1 - \alpha}{\alpha} \deg(i),
   \quad \text{for all $i \in V$}.
\end{equation}
Since $\deg(i) \geq 1$ for all $i \in V$, it suffices to choose
$\epsilon \leq \bar{k}^{-1} \frac{1 - \alpha}{\alpha}$.  For a
(nonuniform) graph, averaging over a regular random walk will quickly
give a positively biased estimate for $\bar{k}$ that can be used to
bound $\epsilon$; note that many nonuniform networks have a modest
average degree, despite the existence of a few extremely high-degree
nodes.

In an implementation of the above sampler one does not of course make
explicit copies of the node sets, but rather uses a state flag that
indicates which set of transition probabilities should be applied. All
the transition probabilities are locally computable at each node $i$,
if the parameters $\epsilon$ and $\alpha$ are given, and the degrees
of both the node $i$ and its neighbours in $\Gamma(i)$ are
accessible. The dependency of $\epsilon'_i$ on the parameter $\alpha$
and the average degree $\bar{k} = 2m/n$ can be resolved by simply
always setting $\epsilon'_i = \displaystyle\frac{\epsilon}{\deg(i)}$,
which implicitly fixes the relationship
\begin{equation}
\frac{\alpha}{1 - \alpha} \cdot \displaystyle\frac{2m}{n} = 1
\Rightarrow \alpha = \displaystyle\frac{1}{\bar{k}+1}.
\end{equation}
This implies by equation~(\ref{eq:resteps}) the condition $\epsilon
\leq 1$, which is a natural restriction on $\epsilon$.  By this choice
of $\epsilon'_i$ we also have $\epsilon'_i \leq \epsilon$ for all $i
\in V$, and may thus set $\delta = \epsilon$, completing the
definition of the transition probabilities on the mixing side.

\begin{figure}[t]
\begin{center}
\begin{picture}(0,0)%
\includegraphics{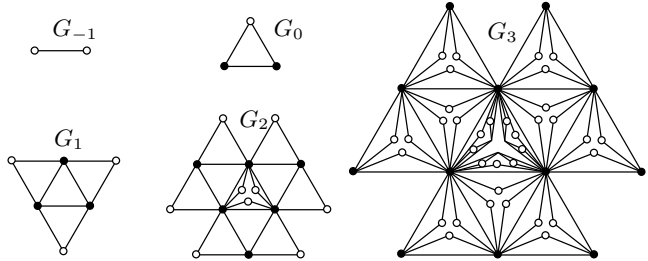}%
\end{picture}%
\setlength{\unitlength}{4144sp}%
\begingroup\makeatletter\ifx\SetFigFont\undefined
\def\x#1#2#3#4#5#6#7\relax{\def\x{#1#2#3#4#5#6}}%
\expandafter\x\fmtname xxxxxx\relax \def\y{splain}%
\ifx\x\y   
\gdef\SetFigFont#1#2#3{%
  \ifnum #1<17\tiny\else \ifnum #1<20\small\else
  \ifnum #1<24\normalsize\else \ifnum #1<29\large\else
  \ifnum #1<34\Large\else \ifnum #1<41\LARGE\else
     \huge\fi\fi\fi\fi\fi\fi
  \csname #3\endcsname}%
\else
\gdef\SetFigFont#1#2#3{\begingroup
  \count@#1\relax \ifnum 25<\count@\count@25\fi
  \def\x{\endgroup\@setsize\SetFigFont{#2pt}}%
  \expandafter\x
    \csname \romannumeral\the\count@ pt\expandafter\endcsname
    \csname @\romannumeral\the\count@ pt\endcsname
  \csname #3\endcsname}%
\fi
\fi\endgroup
\begin{picture}(3826,1547)(28,-716)
\put(311,-39){\makebox(0,0)[lb]{\smash{\SetFigFont{9}{10.8}{rm}{\color[rgb]{0,0,0}$G_1$}%
}}}
\put(301,621){\makebox(0,0)[lb]{\smash{\SetFigFont{9}{10.8}{rm}{\color[rgb]{0,0,0}$G_{-1}$}%
}}}
\put(1625,620){\makebox(0,0)[lb]{\smash{\SetFigFont{9}{10.8}{rm}{\color[rgb]{0,0,0}$G_0$}%
}}}
\put(2904,604){\makebox(0,0)[lb]{\smash{\SetFigFont{9}{10.8}{rm}{\color[rgb]{0,0,0}$G_3$}%
}}}
\put(1420, 78){\makebox(0,0)[lb]{\smash{\SetFigFont{9}{10.8}{rm}{\color[rgb]{0,0,0}$G_2$}%
}}}
\end{picture}

\end{center}
\caption{The DGM pseudo-fractal graphs $G_t$ (adapted from
\protect\cite{pseudofractal}). Newly added nodes are drawn white.}
\label{fig:pseudofract}
\end{figure}

\section{Sampling experiments}
\label{sec:sampling_experiments}

In this section, we report on experiments using the above sampler
construction on both artificial networks with known properties (so
called ``pseudofractal graphs'' of Dorogovtsev, Goltsev, and
Mendes~\cite{pseudofractal}), and scientific collaboration networks of
$n = $ 503 and $n = $ 5,909 mathematicians and computer scientists,
with total number of coauthorships $m =$ 828 and $m = $ 13,510
respectively (subgraphs of the network constructed in~\cite{licsc}).

In the deterministic scale-free network model of
Dorogovtsev, Goltsev, and Mendes~\cite{pseudofractal} (based on \cite{vicsek}),
the initial graph $G_{-1} = (V_{-1}, E_{-1})$ consists of two nodes
$v$ and $w$ and an edge $(v, w)$. At each generation $t \geq 0$
of the generative process, per each edge $(u, v) \in E_{t-1}$,
a new node $w$ is
added together with edges $(u, w)$ and $(v, w)$.
(See Figure~\ref{fig:pseudofract} for an
illustration of the first five generations.)
The resulting graphs $G_t$ have an almost constant average degree of
$\bar{k}_t = 4(1 + 3^{-t})$, yet a power-law distribution of node
degrees according to $n_t(d) = 3^{t+1}d^{-\log_2 3}$.

\begin{figure}[ht!]
\begin{center}
\begin{tabular}{cc }
\includegraphics[width=50mm]{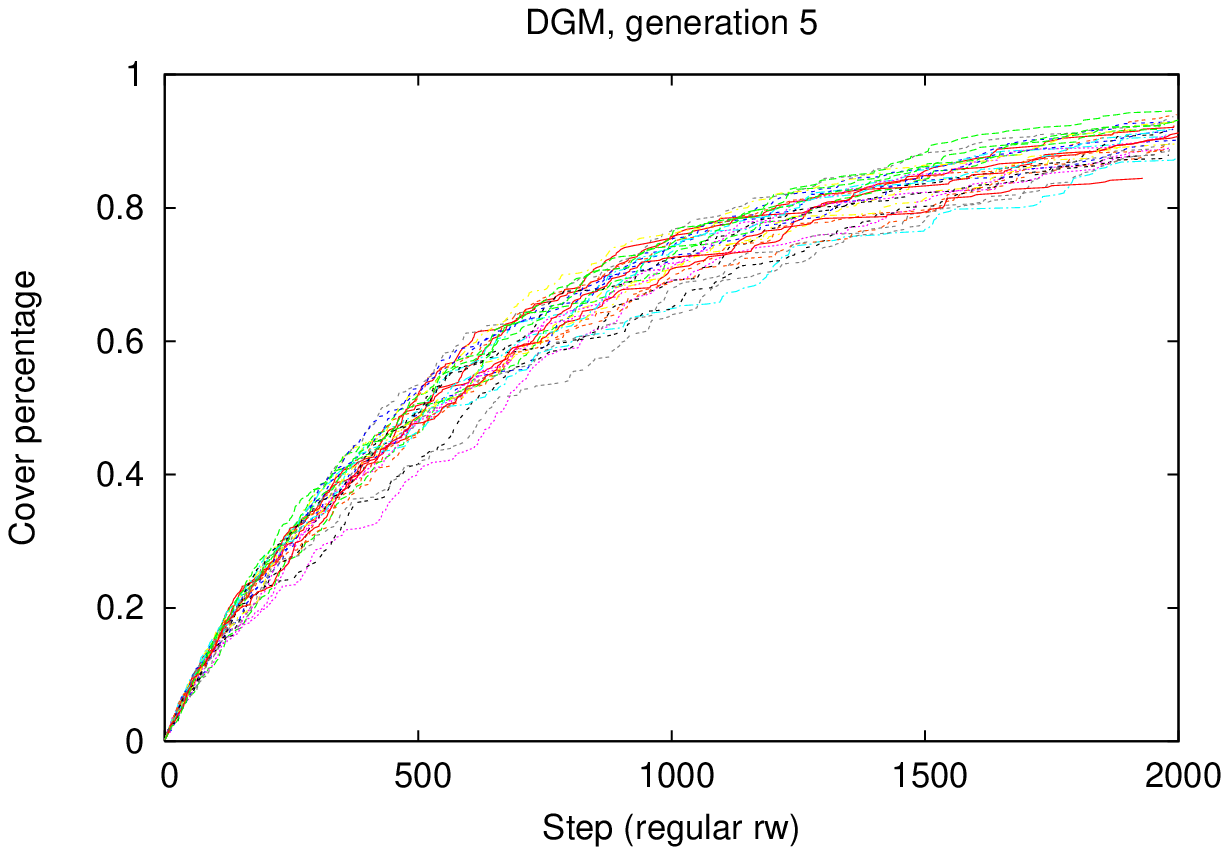} &
\includegraphics[width=50mm]{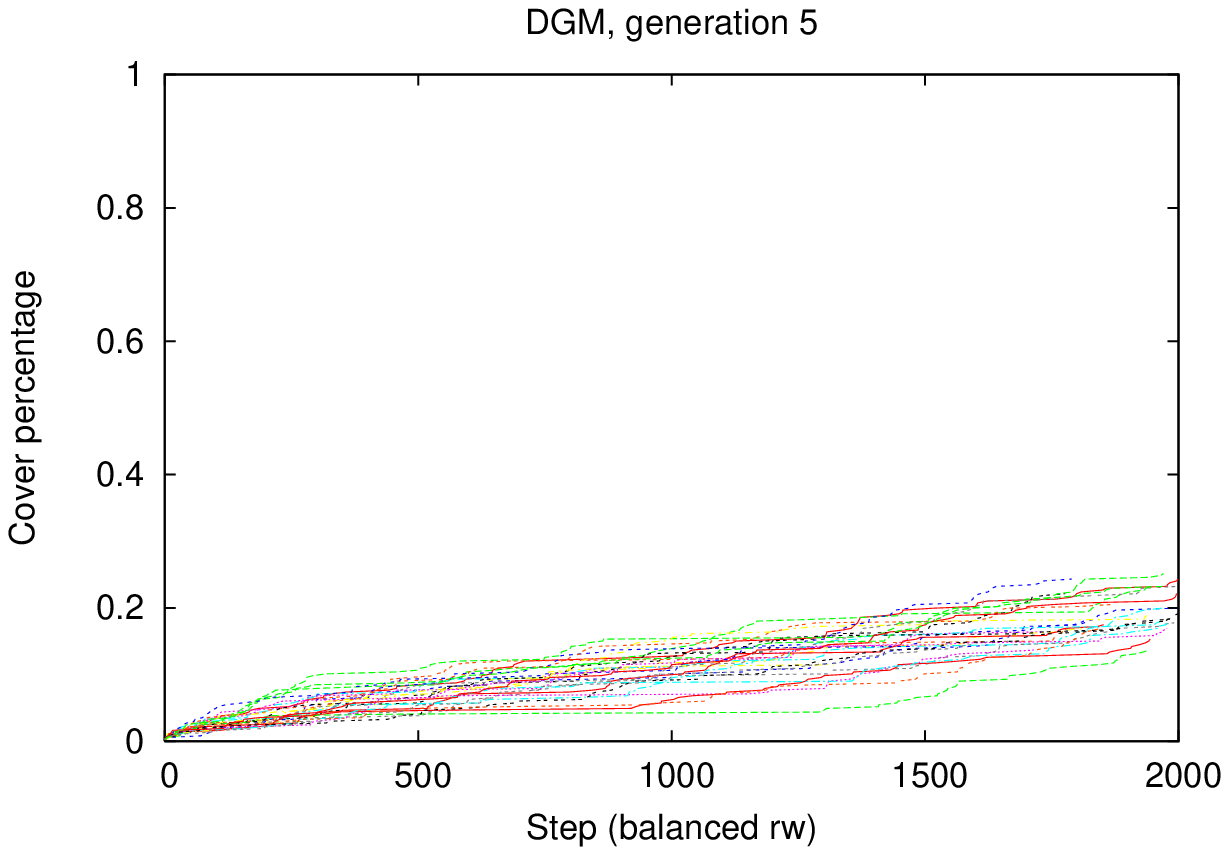} \\
\includegraphics[width=50mm]{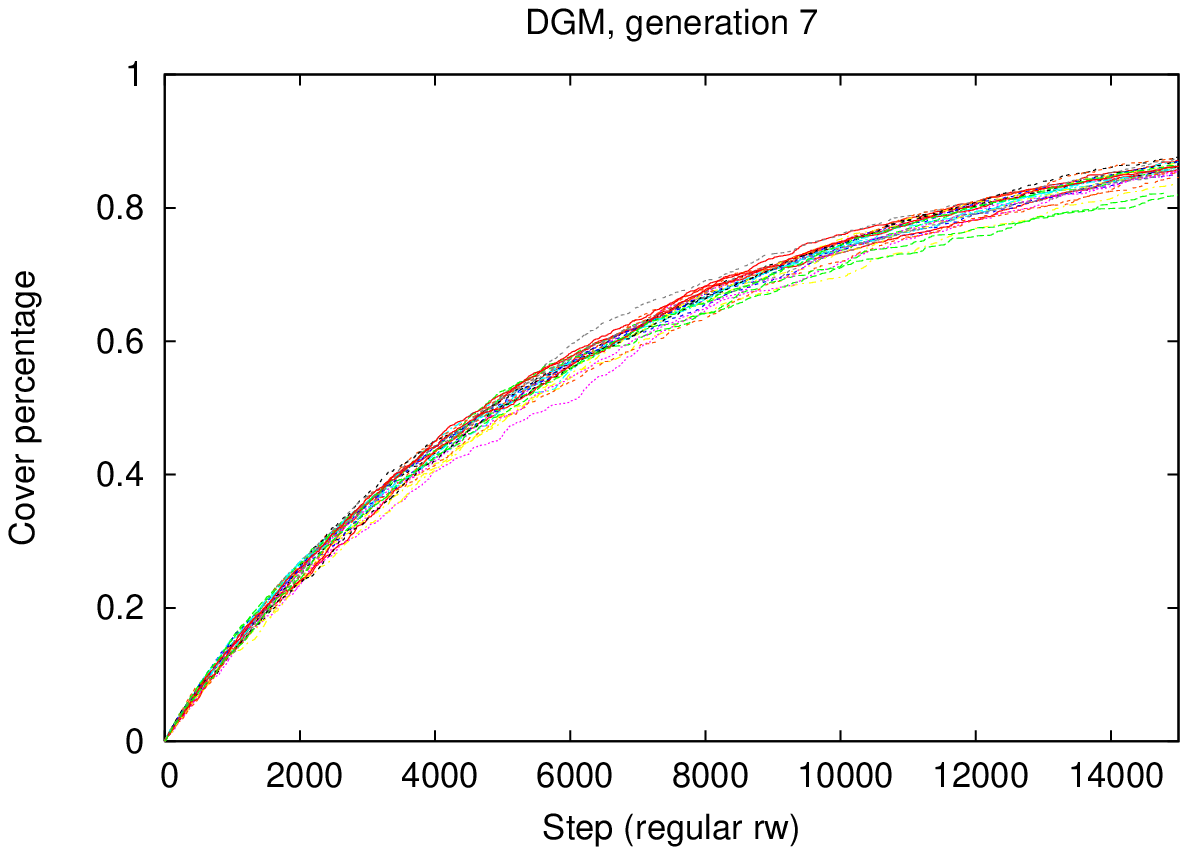} &
\includegraphics[width=50mm]{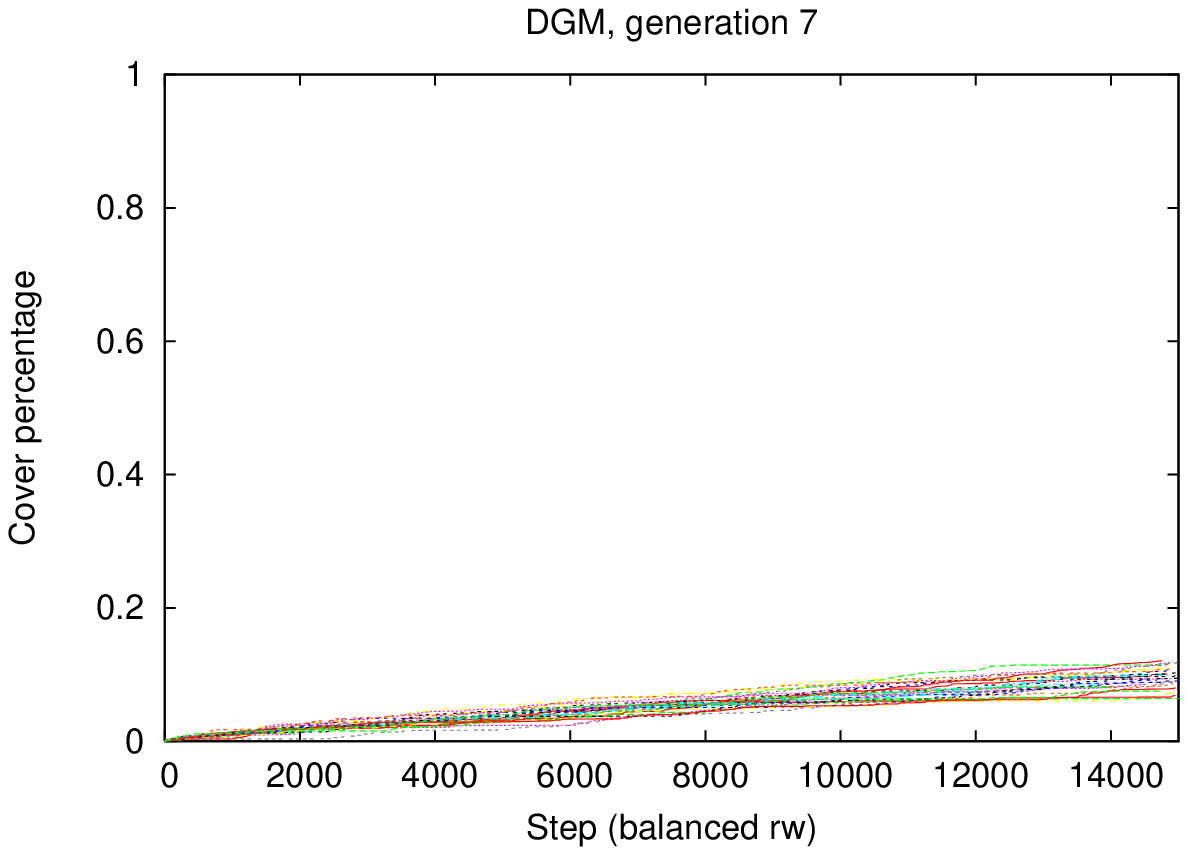} \\
\includegraphics[width=50mm]{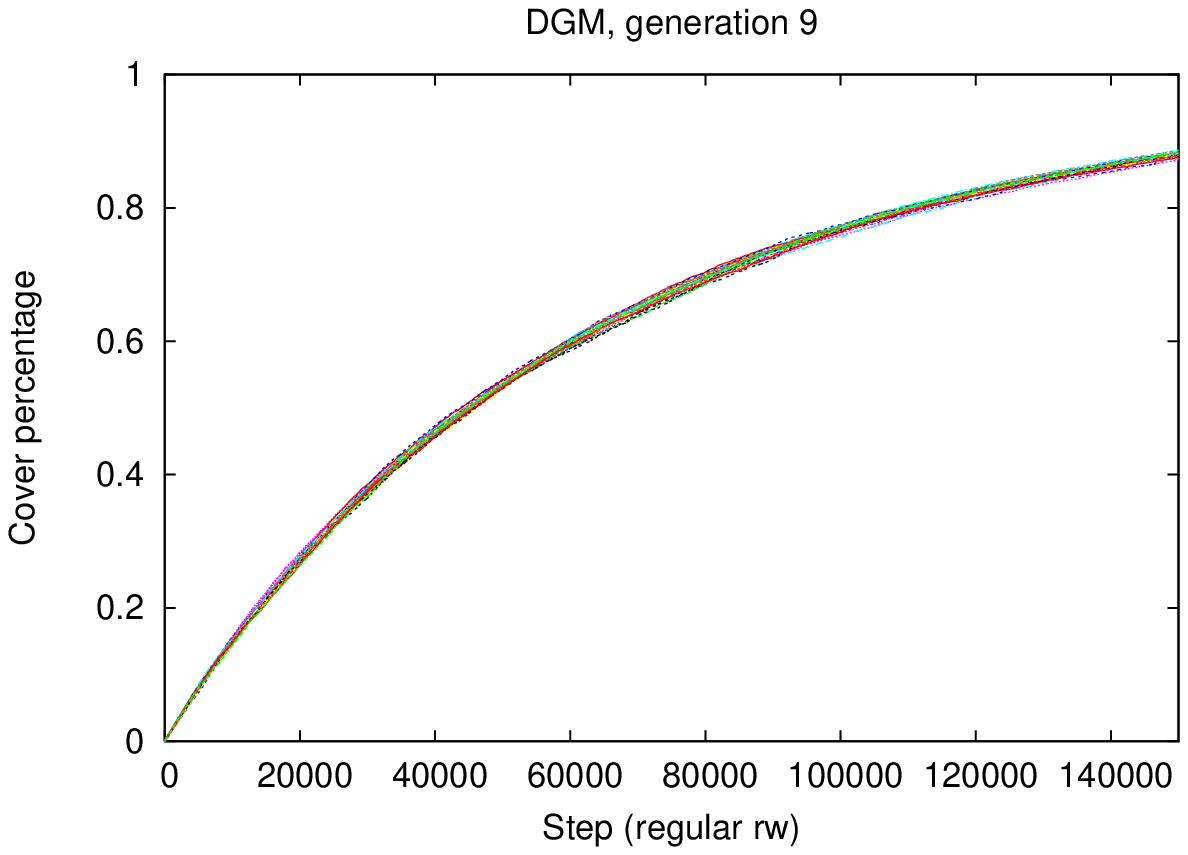} &
\includegraphics[width=50mm]{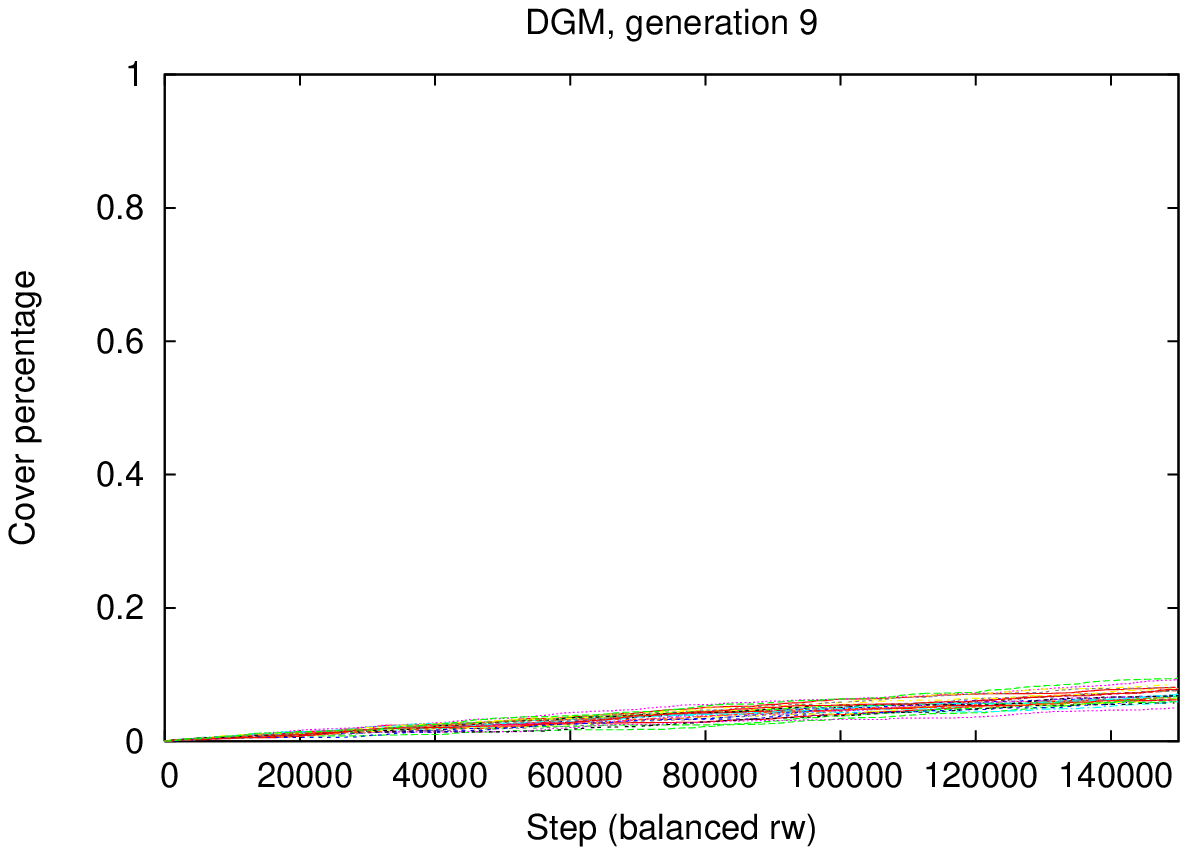} 
\end{tabular}
\end{center}
\caption{The coverage achieved by the regular (top row) and the
degree-balanced walks at each step for DGM graphs of generations
5, 7, and 9. In each plot, 30 independent walks are shown.}
\label{fig:dettricover}
\end{figure}

\begin{figure}
\centerline{\includegraphics[width=50mm]{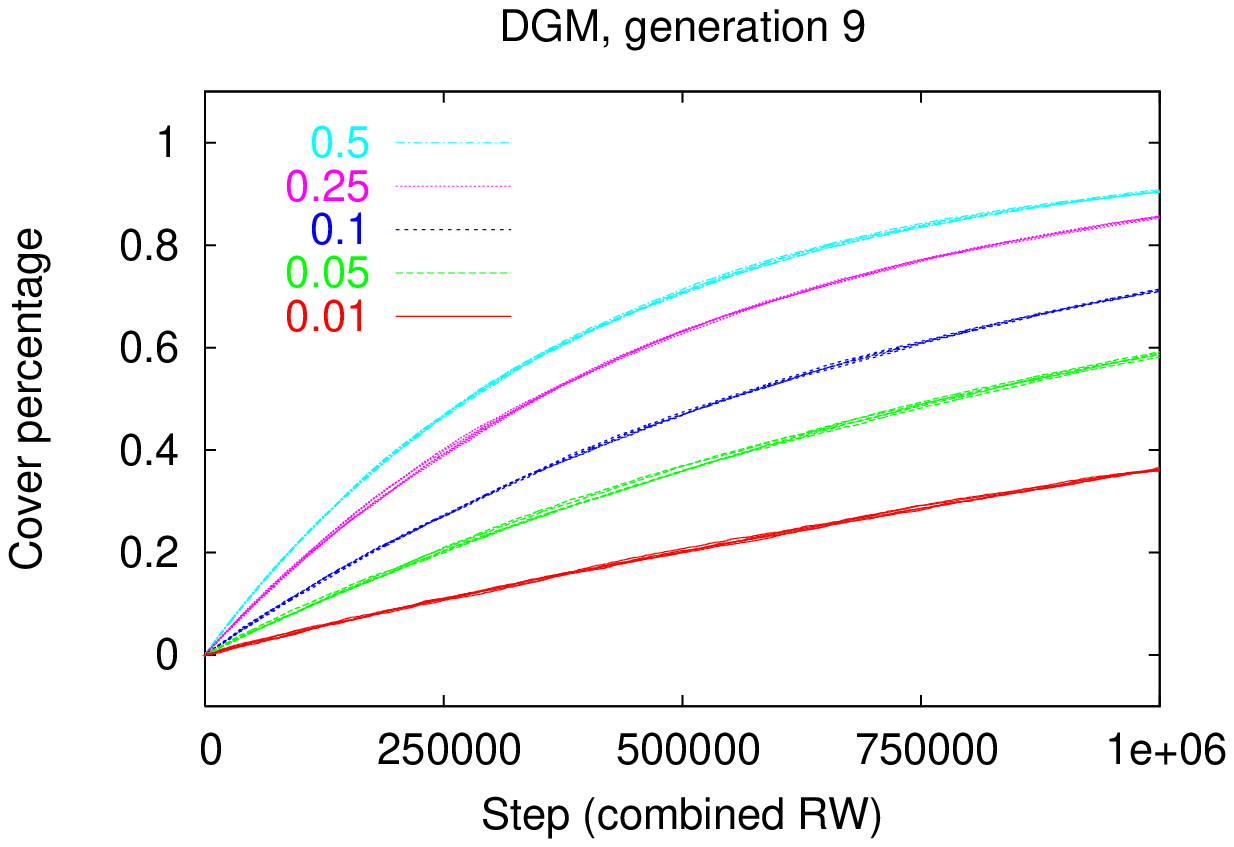}}
\caption{The coverage achieved by the combination random walk on a
ninth generation DGM graph for different values of $\epsilon$.}
\label{fig:dettricombicover}
\end{figure}

As a first indication of the behaviour of various sampling strategies,
Figures~\ref{fig:dettricover} and~\ref{fig:dettricombicover}
present plots of the percentage of graph covered versus
length of the walk, for DGM networks $G_5$, $G_7$ and $G_9$.
Note that the combination walks sample fewer nodes during a walk
of a given length than the others, as it does not record samples
during the mixing phase. The tendency of the degree-balanced method to
unwanted locality is quite evident in Figure~\ref{fig:dettricover}.

In another set of experiments, we estimated the rate of convergence of
the above discussed random walks to their respective stationary
distributions. If $\pi_t$ is the distribution of a random walk after
$t$ steps, and $\pi$ is its stationary distribution, the {\em total
variation distance} between the two is defined
as~\cite{Brem99,Hagg02}:
\begin{equation}
\Delta(t) = \max_{S \subseteq V} | \pi_t(S) - \pi(S) | =
\frac{1}{2} \sum_{i \in V} | \pi_t(i) - \pi(i) |.
\end{equation}
We estimate this quantity by running $k$ independent instantiations of
a given random walk starting from the same start node, and looking at
the state distributions at time $t$ of the instantiations.  For
definiteness, let us consider the case where the stationary
distribution is uniform, with $\pi(i) = 1/n$ for all $i \in
V$.  Denoting by $f_t(i)$ the number of instantiated walks that are
visiting node $i$ at time $t$, a conservative estimate of the total
variation distance at time $t$ can then be computed as~\cite{tvdmcmc}:
\begin{equation}   \label{eq:tvdest}
\Delta_\text{est}(t) = 1 - \sum_{i=1}^n \min \bigg \{
\frac{f_t(i)}{k}, \, \frac{1}{n} \bigg \}.
\end{equation}

\begin{figure}
\begin{center}
\begin{tabular}{cc}
\includegraphics[width=50mm]{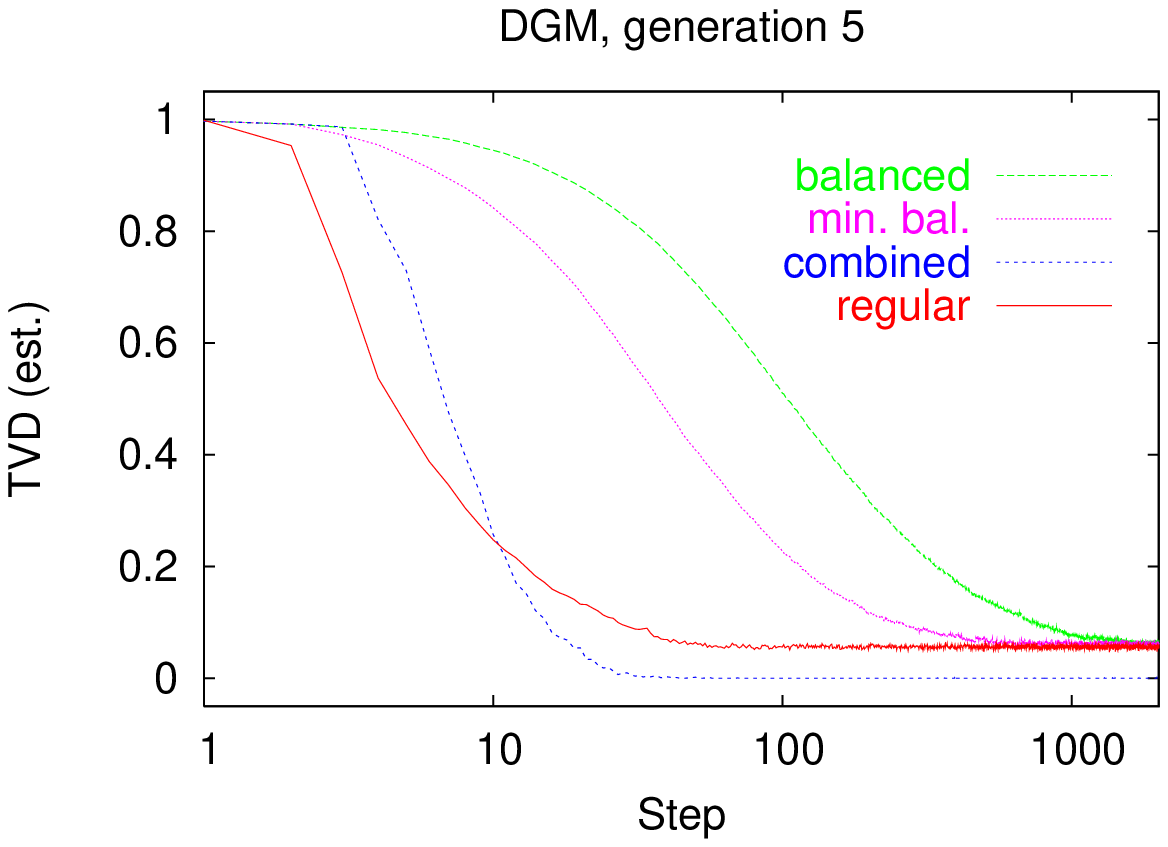} & 
\includegraphics[width=50mm]{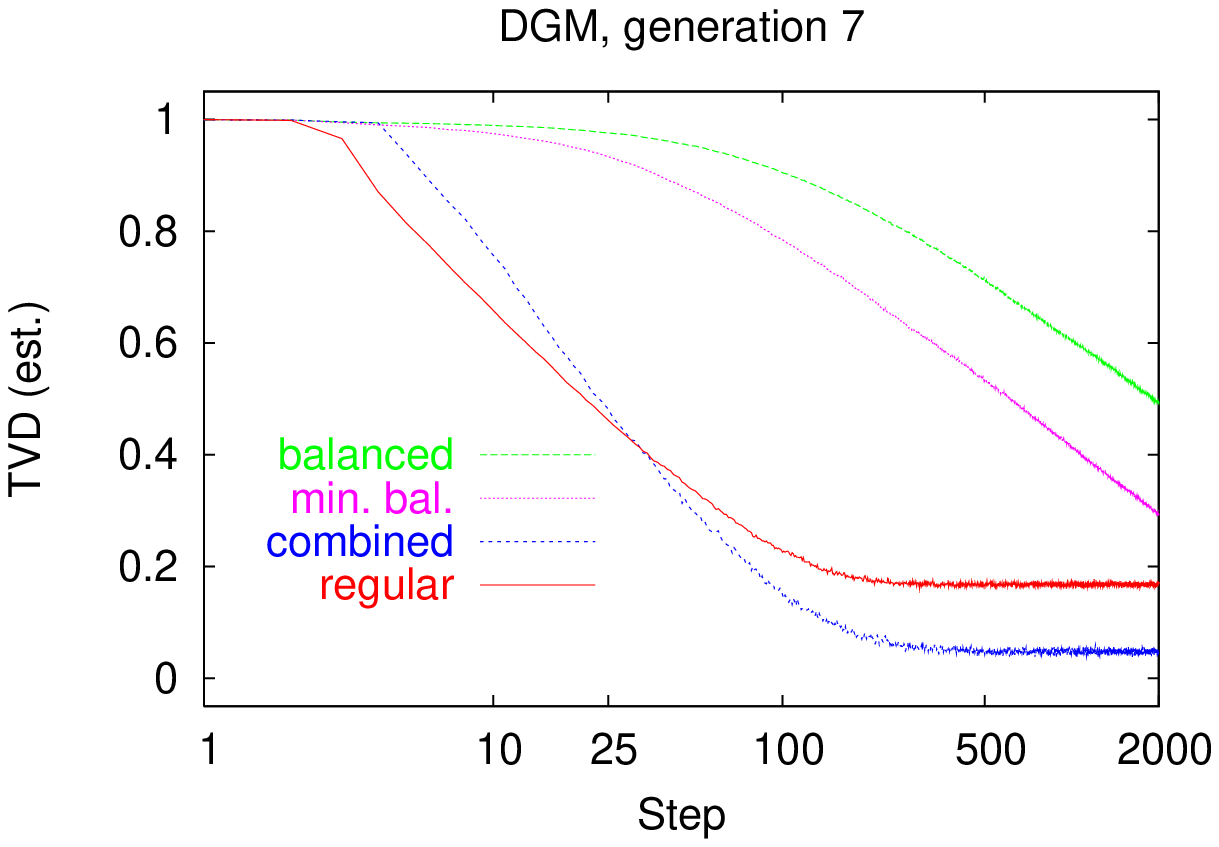} \\
\includegraphics[width=50mm]{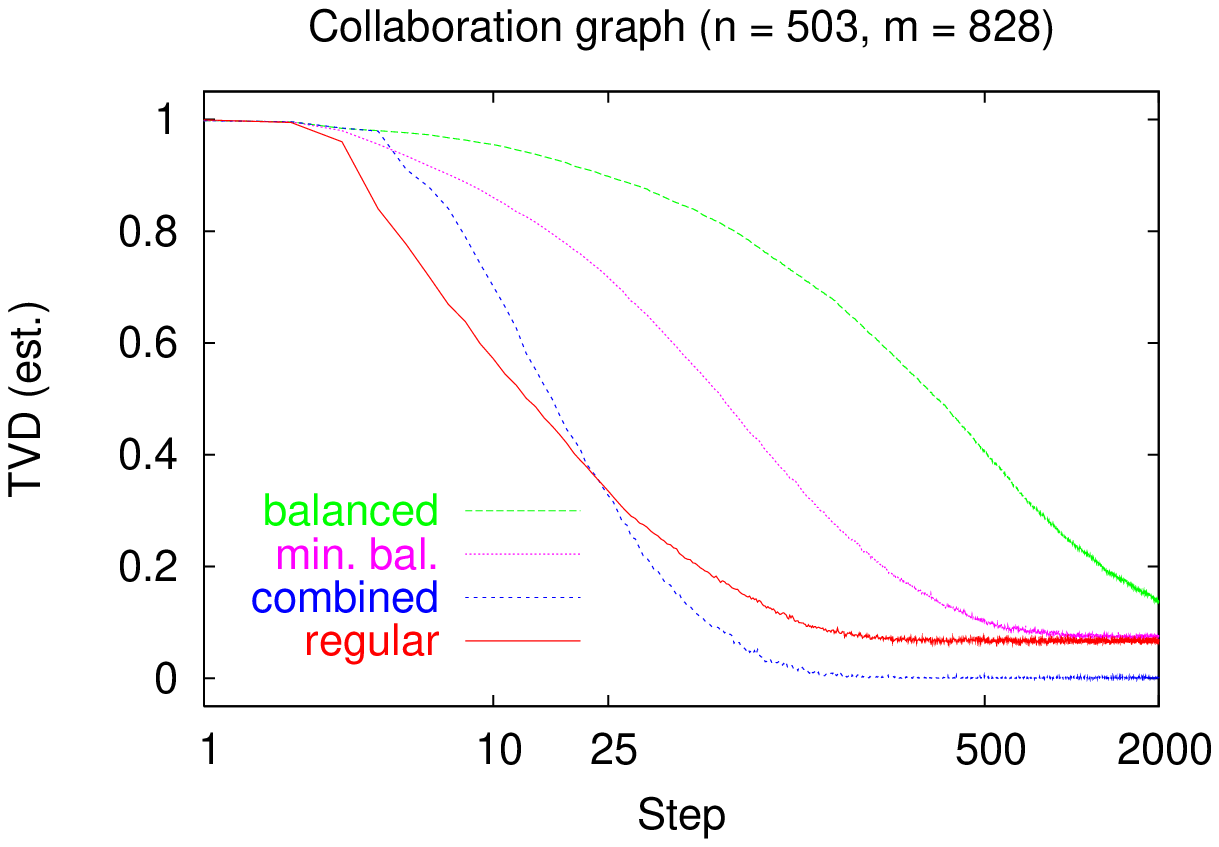} & 
\includegraphics[width=50mm]{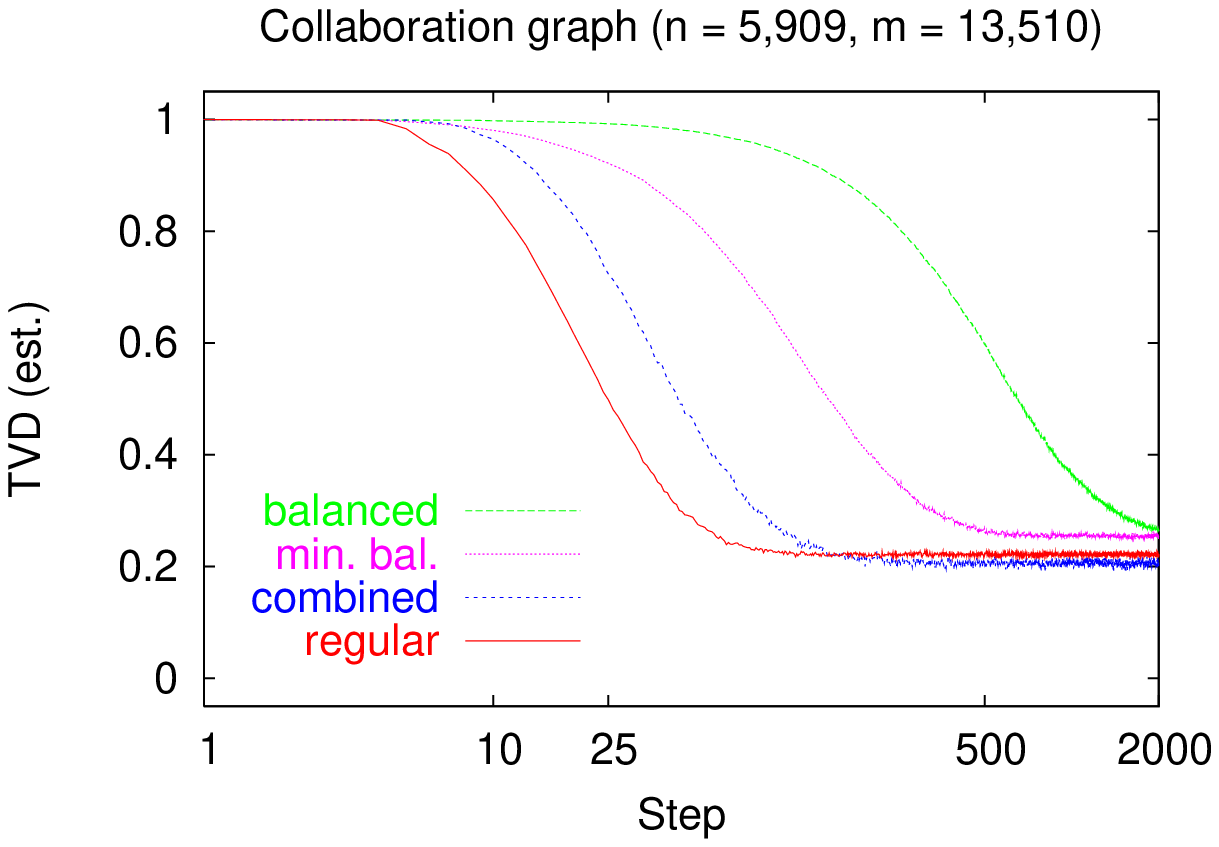}
\end{tabular}
\end{center}
\caption{Values of $\Delta_\text{est}(t)$ for the regular,
degree-balanced, combination, and Metropolis-Hastings
(``minimum-balanced'')~random walks
over a set of 15,000 independent walks in two DGM graphs
and two collaboration graphs, all starting from a fixed node,
initially chosen at random. Note logarithmic scale on the time axis.}
\label{fig:dgmtvd}
\end{figure}

Figure~\ref{fig:dgmtvd} shows the time evolution of these estimates
for the regular, balanced, combination random walks and the
Metropolis-Hastings walk of~\cite{BoDX04} in DGM graphs of generations
five and seven, and for the two collaboration graphs of $n = $ 503 and
$n = $ 5,909 scientists.  The stationary distribution for the regular
walk is taken to be the degree-proportional distribution $\piRW$, and
for the three other walks the uniform distribution $\piID$. For the
combination walk, only those instantiated walks that are on the
sampling side at any given time step are included in computing the
corresponding estimate. The plots illustrate quite graphically
(particularly in the case of the heavy-tailed DGM graphs) that the
convergence behaviour of the combination walk is qualitatively similar
to that of the regular walk, whereas both the pure balanced walk and
the Metropolis-Hastings walk converge noticeably more
slowly.\footnote{ There is some residual small-sample bias in the
estimates; we have computed the size of this effect and will indicate
these calculations in the extended version of this paper.}

\section{Local clustering by approximate Fiedler vectors}

Another key task in the analysis of natural networks is
finding clusters of densely interconnected nodes.
Most of the existing literature on this topic (see~\cite{NeGi04}
for a survey) considers the task
of finding an ideal {\em complete} clustering of
a given graph. This is, however, often unnecessary
and in any case infeasible in the case of really large
networks such as the WWW. (The fastest complete
algorithms can currently deal with networks containing up to
maybe a few millions of nodes~\cite{HKKS03,Newm03,NeGi04}.)
In many cases it would be sufficient to know the relevant
cluster of a given source node, or maybe a group of nodes.
Some recent papers, such as~\cite{Virt03,WuHu04} address
also this more limited goal.

In~\cite{Virt03, licsc}, a parameter-free local clustering quality
measure is optimised using simulated annealing: the computational
effort needed to obtain the cluster of a given source node is quite
modest (and, most importantly, independent of the total size of the
network), and the results seem to be quite robust w.r.t.\ variations
in the annealing process. In~\cite{WuHu04}, the clustering task is
formulated as a problem of determining voltage levels in an electrical
circuit with unit resistances corresponding to the edges of the
original network. The source node is fixed at a high voltage value and
a randomly selected target node at low voltage; an approximate
solution to the Kirchhoff equations is computed by an iteration
scheme, and the eventual cluster of the source node is deemed to
consist of those nodes whose voltages are ``close''~to the high value.
The possibility that the target node is accidentally selected from
within the natural cluster of the source node is decreased by
repeating the experiment some small number of times and determining
cluster membership by majority vote.

This electrical circuit analogue appears to have been first suggested
in~\cite{NeGi04}, where however the aim is to compute a complete
clustering of a given network by considering all possible
source-target pairs, and for each pair solving the Kirchhoff equations
exactly by explicitly inverting the corresponding Laplacian
matrix. (We note that since solutions of the Kirchhoff equations can
be decomposed in terms of the eigenvectors of the circuit graph
Laplacian, this method is a variant of the much-studied spectral
partitioning techniques
\cite{Fied73,Fied75,GkMZ03,GuMi98,KaVV04,PoSL90,SpTe96}.  A
distributed algorithm for spectral analysis, possibly suited for large
networks, is proposed in~\cite{KeMc04}.  A fundamental reference
is~\cite{Chun97}.)

We continue the analogue of representing cluster membership values as
physical potentials, but eliminate the unnatural choice of random
``target'' nodes by basing our model on {\em diffusion in an unbounded
medium} rather than an electrical closed-circuit model. Thus, we fix
the source node $i$ at a constant potential level, which we choose to
be zero, and find an eigenvector $u$ corresponding to the smallest
eigenvalue $\sigma_1$ of the respective {\em Dirichlet matrix}, i.e.\
the Laplacian matrix of the network with row and column $i$
removed~\cite{Chun97,ChEl02}. This eigenvector $u$, called the {\em
(Dirichlet-)Fiedler} vector of the graph, will now (hopefully) assign
potential values $u(j)$ close to 0 for nodes $j$ that are within a
densely interconnected neighbourhood of the source node $i$, and
larger values for nodes that have sparser connections to the
source. (The method obviously generalises to starting from a larger
set of source nodes, if desired.)

Since we wish to develop a local algorithm, and not deal with
the full adjacency matrix of the network, we approach the
computation of the Fiedler vector $u$ via minimising
the Rayleigh quotient~\cite{Chun97,ChEl02}:
\begin{equation}   \label{eq:rayleigh}
\sigma_1 \quad = \quad \inf_u \frac{\sum_{j \sim k} (u(j) - u(k))^2}
                              {\sum_j u(j)^2},
\end{equation}
where the infimum is computed over vectors $u$ satisfying the
Dirichlet boundary condition of having $u(i) = 0$ for the source
node(s). (The notation $j \sim k$ is an abbreviation for $(j,k) \in
E$.) Furthermore, since we are free to normalise our eventual Fiedler
vector to any length we wish, we can constrain the minimisation to
vectors $u$ that satisfy, say, $\|u\|^2_2 = n = |V|$.  Thus, the task
becomes one of finding a vector $u$ that satisfies:
\begin{equation}   \label{eq:fiedler}
u \quad = \quad \argmin \bigg \{\sum_{j \sim k} (u(j) - u(k))^2 \;\mid\;
                          u(i) = 0,\; \|u\|_2^2 = n \bigg \}. 
\end{equation}
We can solve this task approximately by reformulating the
requirement that $\|u\|_2^2 = n$ as a ``soft constraint'' with
weight $c > 0$, and minimising the objective function
\begin{equation}   \label{eq:soft_fiedler}
f(u) \quad = \quad \frac{1}{2} \sum_{j \sim k} (u(j) - u(k))^2 +
                   \frac{c}{2} \cdot \big (n - \sum_j u(j)^2\big)
\end{equation}
by gradient descent. Since the partial derivatives of $f$ have
the simple form
\begin{equation}   \label{eq:soft_partials}
\frac{\partial f}{\partial u(j)} \quad = \quad
   - \sum_{k \sim j} u(k) + (\deg(j) - c) \cdot u(j),
\end{equation}
the descent step can be computed locally at each node, based on
information about the $u$-estimates at the node itself and its
neighbours:
\begin{equation}   \label{eq:grad_desc}
\tilde{u}_{t+1}(j) \quad = \quad \tilde{u}_t(j) + \delta \cdot
   \big(\sum_{k \sim j} \tilde{u}(k) - (\deg(j) - c) 
\cdot \tilde{u}(j)\big),
\end{equation}
where $\delta > 0$ is a parameter determining the speed of the descent.
Assuming that the natural cluster of node $i$ is
small compared to the size of the full network, the normalisation
$\|u\|_2^2 = n$ entails that most nodes $j$ in the network will
have $u(j) \approx 1$. Thus the descent iterations~(\ref{eq:grad_desc})
can be started from an initial vector $\tilde{u}_0$ that has
$\tilde{u}_0(i) = 0$ for the source node $i$
and $\tilde{u}_0(k) = 1$ for all $k \neq i$.
The estimates need then to be updated at time $t > 0$ only for
those nodes $j$ that have neighbours $k \sim j$ such that
$\tilde{u}_{t-1}(k) < 1$.

\begin{figure}
\begin{center}
\begin{tabular}{ccc}
\includegraphics[width=38mm]{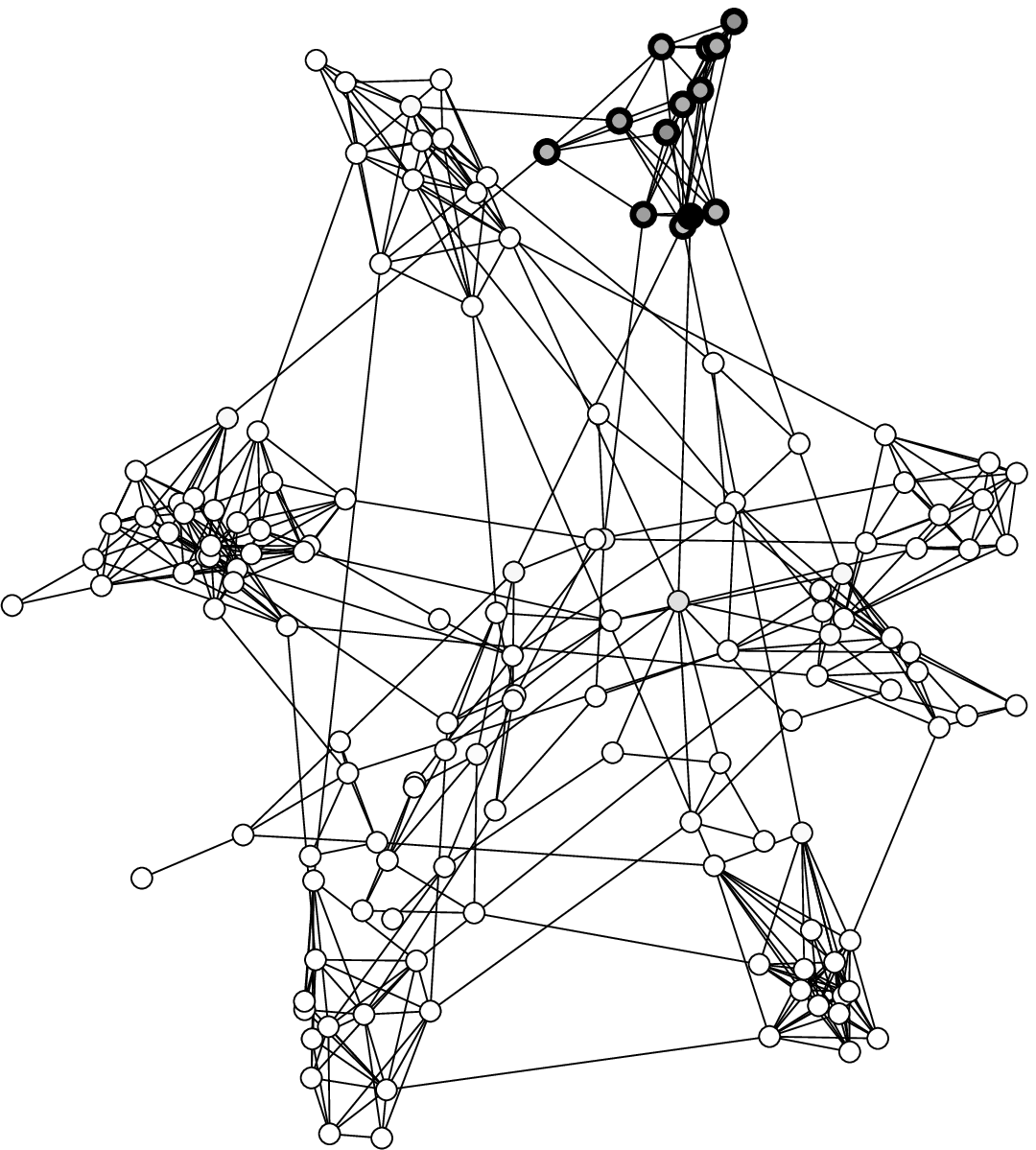} &
\includegraphics[width=38mm]{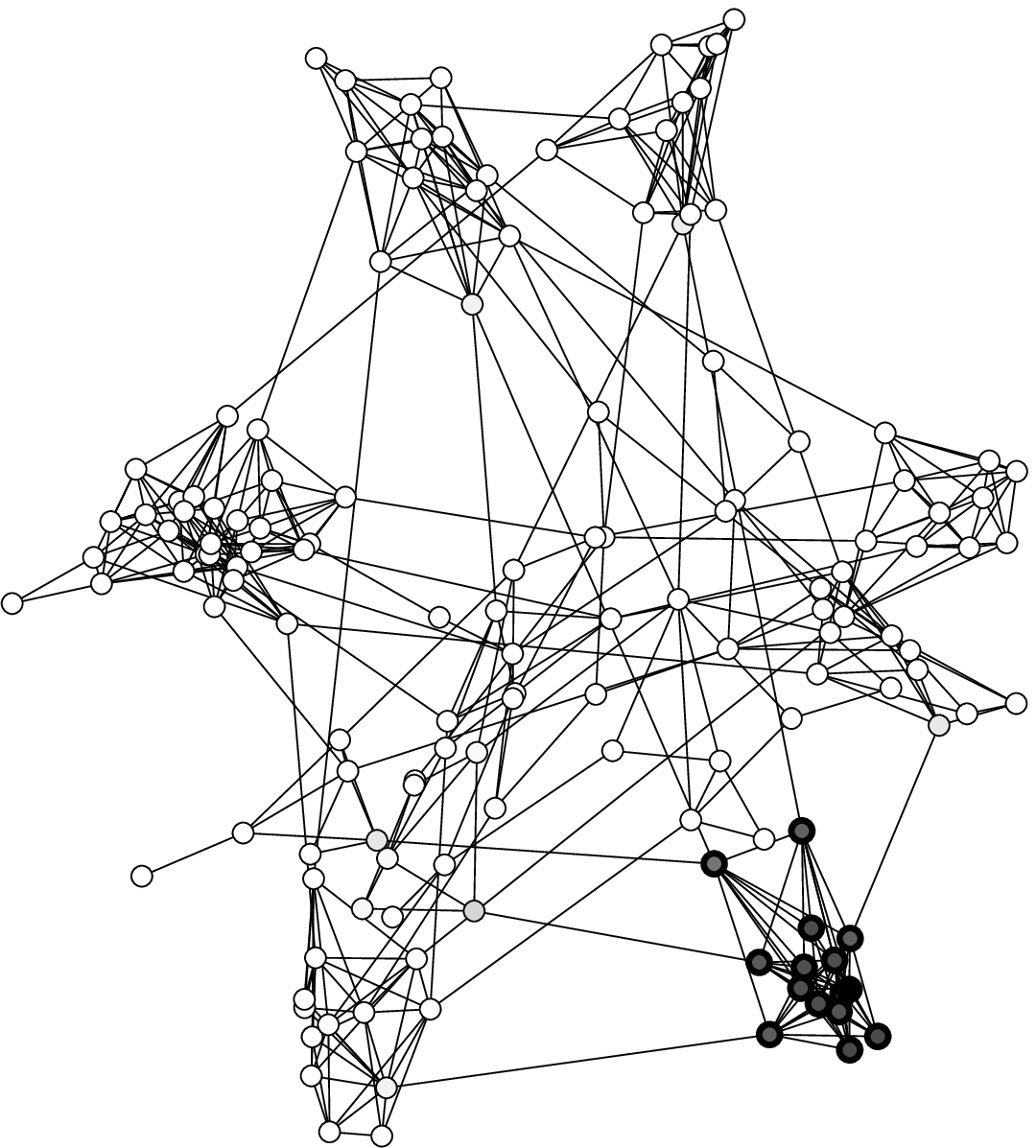} &
\includegraphics[width=38mm]{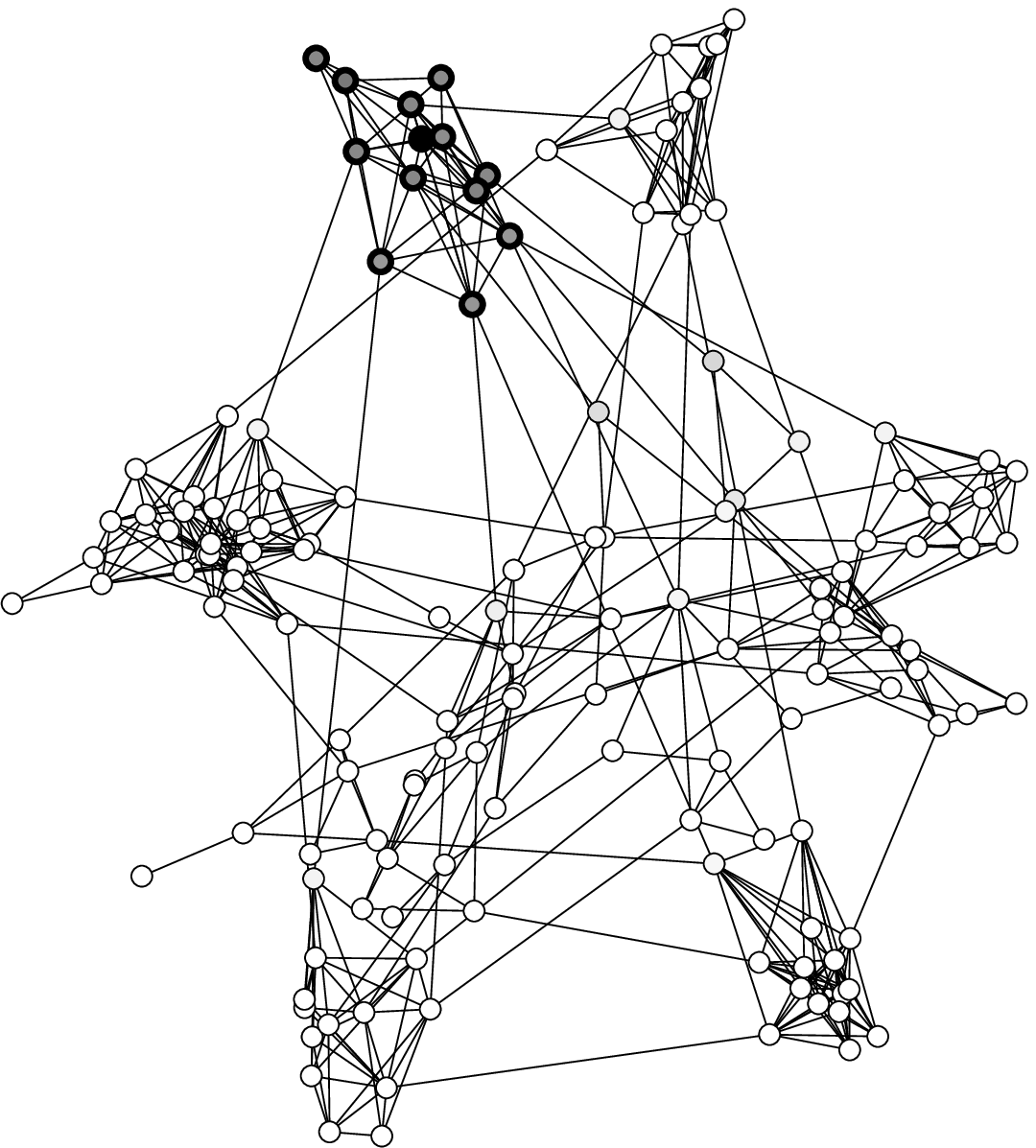}
\end{tabular}
\end{center}
\caption{Three local Fiedler clusters in a caveman graph of 138 nodes.}
\label{fig:cave_clusters}
\end{figure}

Figure~\ref{fig:cave_clusters} represents the results of such
approximate Fiedler vector calculations in the case of a slightly
randomised ``caveman'' network of 138 nodes, starting from three
different source nodes. For visual effect, the nodes are colour-coded
so that dark colours correspond to small approximated Fiedler
potential values, with the source node in each case coloured
black. The parameter values used in this case were $c = 0.1$, $\delta
= 0.05$.

Visually, the clusters in e.g.\ Figure~\ref{fig:cave_clusters} look
reasonable; in practice, however, we need to determine the cluster
boundaries automatically. One possibility would be to threshold the
potentials as in~\cite{WuHu04}, but we prefer not to introduce any
additional instance-specific parameters to the algorithm. A natural
alternative is to find a set of nodes $S$ that contains the source
node $i$ and minimises some {\em weighted Cheeger ratio}~\cite[p.\
35]{Chun97}:
\begin{equation}   \label{eq:cheeger}
h_w(S) \quad = \quad
   \frac{\sum_{j \in S} \sum_{k \sim j, k \not\in S} w(j,k)}
        {\sum_{j \in S} \sum_{k \sim j} w(j,k)},
\end{equation}
where $w(j,k)$ is an appropriate nonnegative edge weight function.  In
our experiments, edge weights determined as $w(j,k) = (|u(j) -
u(k)|)^{-1}$ seem to lead to natural clusters in different types of
networks, and are also intuitively appealing.  In
Figure~\ref{fig:cave_clusters}, we have indicated the nodes selected
by this heuristic as belonging to each cluster by circles with thick
boundaries. The minimisation of the cluster cost
function~(\ref{eq:cheeger}) was here performed by a local simulated
annealing process similar to the one used in~\cite{Virt03, licsc}.

\section{Conclusions and Further Work}

In this paper we presented two methods to help analyse properties of
large nonuniform graphs: a uniform sampling construction and a local
method for clustering based on approximate Fiedler vectors. According
to our experiments, both approaches are well-behaving and conform to
the intuition that arises from their analytical properties. 

As future work, we will look into more general constructions for
rapidly mixing uniform MCMC samplers; one direction might be to
combine the regular random walk with alternative slow uniform
samplers, such as those of~\cite{BoDX04}. Accuracy of the estimates of
natural network network characteristics based on our pseudo-uniform
samples should also be assessed. Both the sampling and the clustering
algorithms should also be extended to work on directed graphs, in
order to deal with interesting natural networks such as the WWW.

\subsection*{Acknowledgments}

We thank most appreciatively Mr.\ Kosti Rytkönen for providing us with
his automatic graph drawing tool used to produce the diagrams in
Figure~\ref{fig:cave_clusters}.

{\small

\bibliographystyle{abbrv}
\bibliography{thesis,sampclus}

}

\end{document}